\documentclass[conference]{IEEEtran}

\ifCLASSINFOpdf
  
\else
  
\fi

\usepackage{graphicx}
\usepackage{float}
\usepackage{amssymb}
\usepackage{amsmath}
\usepackage [autostyle, english = american]{csquotes}
\MakeOuterQuote{"}
\usepackage[bookmarks=false]{hyperref}

\usepackage{subcaption}

\begin{document}

\title{DRL-based Distributed Resource Allocation for Edge Computing in Cell-Free Massive MIMO Network}

%\title{Multi-Agent Reinforcement Learning- Resource Allocation over Cell-Free Massive %MIMO-enabled Mobile Edge Network}

\author{\IEEEauthorblockN{Fitsum Debebe Tilahun\IEEEauthorrefmark{1},
Ameha Tsegaye Abebe\IEEEauthorrefmark{2}, and
Chung G. Kang\IEEEauthorrefmark{1}}
\IEEEauthorblockA{\IEEEauthorrefmark{1}School of Electrical Engineering, Korea University, Seoul, Republic of Korea\\
\IEEEauthorrefmark{2}Samsung Research, Seoul, Republic of Korea\\
Email: \IEEEauthorrefmark{1}\{fitsum\_debebe, ccgkang\}@korea.ac.kr,
\IEEEauthorrefmark{2}amehat@samsung.com}}

%\author{\IEEEauthorblockN{Fitsum Debebe Tilahun, Ameha Tsegaye Abebe, and Chung G. Kang}
%\IEEEauthorblockA{School of Electrical Engineering, Korea University\\
%Seoul, Korea, Republic of\\}
%Email:\{fitsum\_debebe, ameha\_tsegaye, ccgkang\}@korea.ac.kr
%}

\maketitle

\begin{abstract}
In this paper, with the aim of addressing the stringent computing and quality-of-service (QoS) requirements of recently introduced advanced multimedia services, we consider a cell-free massive MIMO-enabled mobile edge network. In particular, benefited from the reliable cell-free links to offload intensive computation to the edge server, resource-constrained end-users can augment on-board (local) processing with edge computing. To this end, we formulate a joint communication and computing resource allocation (JCCRA) problem to minimize the total energy consumption of the users, while meeting the respective user-specific deadlines. To tackle the problem, we propose a fully distributed solution approach based on cooperative multi-agent reinforcement learning framework, wherein each user is implemented as a learning agent to make joint resource allocation relying on local information only. The simulation results demonstrate that the performance of the proposed distributed approach outperforms the heuristic baselines, converging to a centralized target benchmark, without resorting to large overhead. Moreover, we showed that the proposed algorithm has performed significantly better in cell-free system as compared with the cellular MEC systems, e.g., a small cell-based MEC system. 
\end{abstract}


\IEEEpeerreviewmaketitle

\section{introduction}
Over the past few years, there has been a surge in computation hungry, and yet highly delay-intolerant multimedia applications. To support these applications in resource-limited end-user devices, it is essential to augment on-board (local) processing by offloading part of the intensive computation to more powerful computing platforms at the network edge. To this end, the ubiquitous radio and computing resources at the mobile edge network should be jointly optimized in a way to improve certain system objectives such as execution delay and energy consumption. In this regard, a plethora of joint communication and computing resource allocation (JCCRA) schemes based on conventional optimization methods have been investigated. However, most of these approaches either assume accurate knowledge of network-wide information, which is difficult to obtain in practice, or are limited to quasi-static systems, such as deterministic task size and time-invariant channel conditions. Thus, their application to support time-sensitive applications in dynamic mobile edge network might be critically limited. 

Driven by the recent advances in machine learning and deep reinforcement learning (DRL), several learning-based schemes have been proposed to provide flexible joint resource allocation. For instance, deep Q-network (DQN) is employed in [1, 2] for designing joint offloading and resource allocation decisions, while [3] presents a deep deterministic policy gradient (DDPG)-based scheme in heterogeneous networks with multiple users. However, these schemes rely on central entities to make joint resource allocation decision, suffering from considerable signaling overhead and scalability issues. [4] investigates distributed JCCRA problem in a single-cell mobile edge computing (MEC) system, however, the adopted computation model is not applicable for tasks with extreme reliability and ultra-low hard deadline requirements. We note that the large amount of effort on optimizing joint resource allocation in MEC networks focuses on cellular system. However, as we move towards the next decade, the current cellular MEC systems may not keep up with the diverse and more stringent requirements of the envisaged advanced applications, such as holographic displays, multi-sensory extended reality, and others. 

In light of the recently proposed cell-free massive MIMO architecture, envisioned for beyond-5G and 6G networks, however, there are only limited works. Notably, cell-free architecture can provide sufficiently fast and reliable access links without cell-edge by serving a relatively small number of users from several geographically distributed access points (APs) [5]. It, therefore, opens up a new opportunity to support energy-efficient and consistently-low latency computational task offloading, in contrast to cellular MEC systems. [6] presents several performance analyses in a cell-free MEC system considering coverage radius of the APs, while the issues of active user detection and channel estimations are discussed in [7].  In this paper, we consider a JCCRA problem that minimizes total energy consumption of the users while meeting the user-specific application deadlines by jointly optimizing the allocation of local processor clock speed and uplink transmission power for each user.  We then present a fully distributed solution approach based on cooperative multi-agent reinforcement learning, alleviating the signaling and communication overheads associated with centralized implementations. In sharp contrast to our work, both [6] and [7] do not deal with the JCCRA problem. To the best of our knowledge, this is the very first attempt to solve JCCRA problem in a fully distributed fashion for cell-free network-enabled mobile edge network. The fully distributed and intelligent JCCRA in our framework combined with the reliable access links of the cell-free massive MIMO architecture can be   a promising means to handle the stringent requirements of the newly introduced multimedia services. 

The rest of the paper is organized as follows: In Section II, we discuss the system model and problem formulation for the JCCRA framework. Section III presents the proposed distributed JCCRA scheme based on cooperative multi-agent reinforcement learning (MARL) framework. Simulation results are presented in Section IV, followed by some concluding remarks in Section V. 
\section{system model and problem formulation }
\subsection{User-centric Cell-free Massive MIMO: Overview}
We consider a cell-free massive MIMO-enabled mobile edge network, depicted in Fig. 1, where a central processing unit (CPU) with an edge server of finite capacity ${f^{CPU}}$ (in cycles per second) is connected to $M$ single-antenna access points (APs) via ideal fronthaul links. Furthermore, $M$ APs serve $K$ single-antenna users, such that $M\gg K$, entailing to the widely known system model for cell-free massive MIMO [5]. Let  ${{\cal M}} = \left\{ {1,2,...,M} \right\}$ and  ${{\cal K}} = \left\{ {1,2,...,K} \right\}$ denote sets of AP and user indices, respectively. Besides, each user $k \in {{\cal K}}$ is also equipped with limited local processor with maximum capability $f_{k}^{\max }$ (in cycle per second).

Let the channel between $m$-th AP and $k$-th user is given as ${{g}_{mk}}=\beta _{mk}^{{}^{1}/{}_{2}}{{h}_{mk}}$ where ${{\beta }_{mk}}$ is a large-scale channel gain coefficient that models the effects of path loss and shadowing while ${h_{mk}} \sim {{\cal C}{\cal N}}\left( {0,1} \right)$ represents small-scale channel fading. Let ${{\tau }_{c}}$ denote a channel coherence period in which ${{h}_{mk}}$ remains the same. The coherence period ${{\tau }_{c}}$ is divided into pilot transmission interval of ${{\tau }_{p}}$ samples and uplink data transmission interval of (${{\tau }_{c}}-{{\tau }_{p}}$) samples for offloading computation to the edge server in the CPU.

\begin{figure}[!t]       %   !t  -> to take it to the top
\centering
\includegraphics[width=2.6in,height=2.3in]{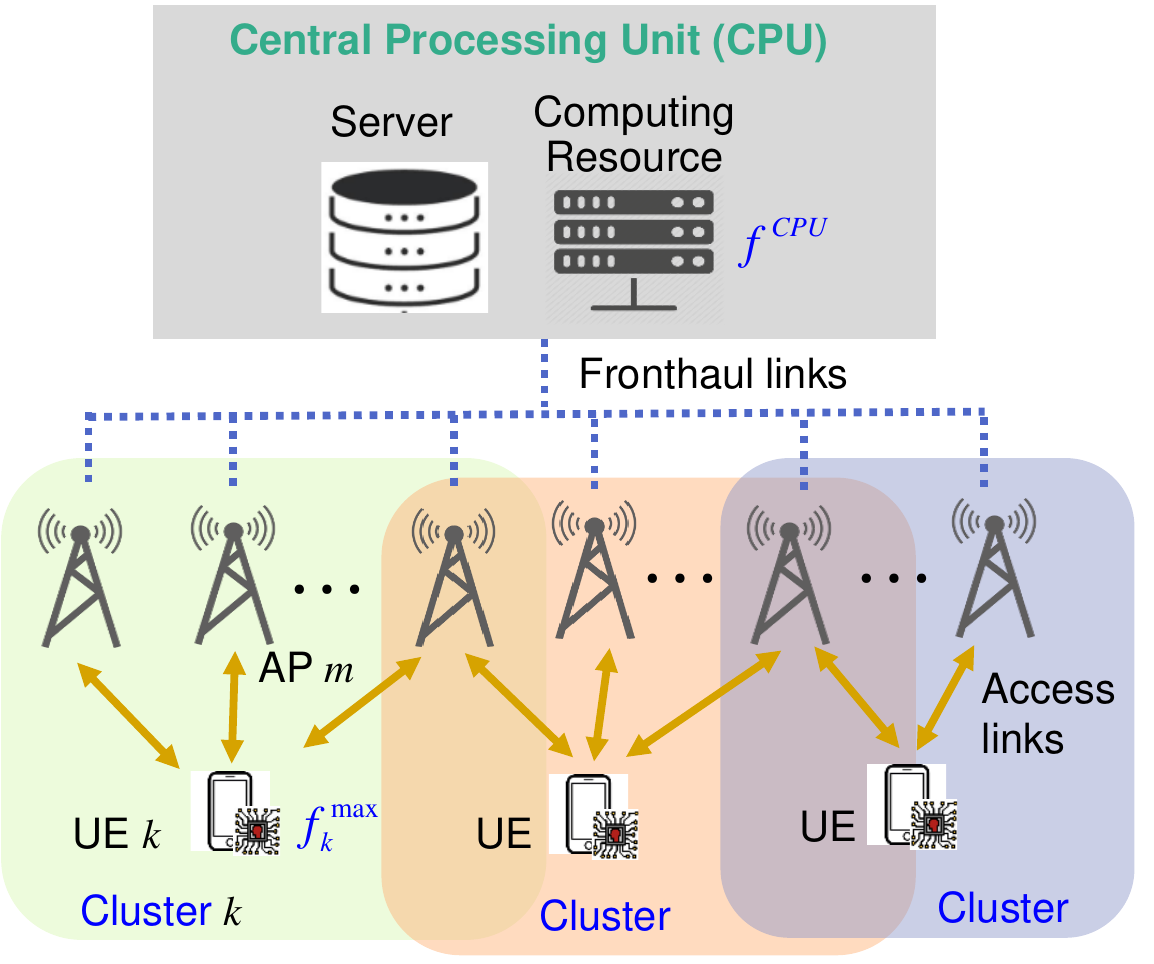}
% where an .eps filename suffix will be assumed under latex, 
% and a .pdf suffix will be assumed for pdflatex; or what has been declared
% via \DeclareGraphicsExtensions.
\caption{User-centric cell-free massive MIMO-enabled mobile edge network:  \emph{Illustrative system model} }
\vspace{-3mm}
\end{figure}
At the beginning of each coherence time, each user $k \in {{\cal K}}$ simultaneously transmit a pilot sequence ${{\mathbf{\psi }}_{k}}\in {{\mathsf{\mathbb{C}}}^{{{\tau }_{p}}\times \,1}}$, such that ${{\left\| {{\mathbf{\psi }}_{k}} \right\|}^{2}}=1$,  which is used to estimate the respective channels ${{\hat{g}}_{mk}}$ at each AP $m\in{{\cal M}}$. Assuming pairwise orthogonal sequences $\left\{ {{\mathbf{\psi }}_{1}},{{\mathbf{\psi }}_{2}},.....,{{\mathbf{\psi }}_{K}} \right\}$, i.e., ${{\tau }_{p}}=K$, we ignore pilot contamination problem in the current model. Then, the received pilot vector at the $m$-th AP, $\mathbf{y}_{m}^{p}\in {{\mathsf{\mathbb{C}}}^{{{\tau }_{p}}\times \,1}}$, can be represented as 
\begin{equation}
{\bf{y}}_m^p = \sqrt {{\tau _p}} \sum\limits_{k = 1}^K {\sqrt {q_k^p} {g_{mk}}{{\bf{\psi }}_k}}  + {\bf{\omega }}_m^p, 
\end{equation}
where $q_k^p$ is the pilot transmit power, and $\mathbf{\omega }_{m}^{p}$ is a $\,{{\tau }_{p}}$-dimensional additive noise vector with independent and identically distributed entries of ${\bf{\omega }}_m^p \sim {{\cal C}{\cal N}}\left( {0,\sigma _{m}^{2}} \right)$. Based on the received vector $\mathbf{y}_{m}^{p}$, the least-square (LS) channel estimate ${{\hat{g}}_{mk}}$ can be expressed as:
\begin{equation}
{\hat g_{mk}} = \frac{1}{{\sqrt {{\tau _p}p_k^p} }}{\bf{\psi }}_k^H{\bf{y}}_m^p.
\end{equation}
The channel estimates are used to decode the uplink transmitted data of the users. After pilot transmission, the users transmit offloaded data to the APs. Let  ${{x}_{k}}$ denote the uplink data of user $k$, with transmit power of ${{p}_{k}}$, which is determined as  ${{p}_{k}}={{\eta }_{k}}\,p_{k}^{\max }$, where ${{\eta }_{k}}$, and $p_{k}^{\max }$ represent power control coefficient and maximum uplink power of the $k$-th user, respectively. Then, the received signal $y_{m}^{u}$ at the $m$-th AP is represented as
\begin{equation}
y_m^u = \sum\limits_{k = 1}^K {\sqrt {{p_k}} {g_{mk}}{x_k}}  + {\omega _m}.
\end{equation}
It is important to ensure the scalability of cell-free massive MIMO system, in terms of complexity for pilot detection and data processing. To this end, we limit the number of APs that serve the $k$-th user to ${N_k} < M$, and form a user-centric cluster of APs ${{{\cal C}}_k} \subset \left\{ {A{P_1},A{P_2}, \cdots ,A{P_{N_k}}} \right\}$, by including all APs with the largest ${{\beta }_{mk}}$ to ${{{\cal C}}_k}$ until the maximum allowable cluster size is reached, i.e., ${N_k} = {C^{\max }}$. Thus, the transmitted data by user $k$ is only decoded by the APs in ${{{\cal C}}_k}$. Then, after performing maximum ratio combining (MRC) at each AP $m \in {{{\cal C}}_k}$ the quantity $\hat{g}_{mk}^{*}y_{m}^{u}$ is transmitted to the CPU via a fronthaul link. The received soft estimates are combined at the CPU to decode the data transmitted by user $k$ as follows:  
\begin{equation}
{\hat x_k} = \sum\limits_{m = 1}^{N_k} {\hat g_{mk}^ * y_m^u}.
\end{equation}
Then, the uplink signal-to interference and noise ratio (SINR) ${\gamma _k}$  for user $k$ can be expressed as:
\begin{equation}
{\gamma _k} = \frac{{{p_k}{{\left| {\sum\limits_{m \in {{{\cal C}}_k}} {\hat g_{mk}^ * {g_{mk}}} } \right|}^2}}}{{\sum\limits_{k' \ne k} {{p_{k'}}} {{\left| {\sum\limits_{m \in {{{\cal C}}_k}} {\hat g_{mk}^ * {g_{mk'}}} } \right|}^2} + \sigma _m^2\sum\limits_{m \in {{{\cal C}}_k}} {{{\left| {{{\hat g}_{mk}}} \right|}^2}} }}
\end{equation}
%\begin{equation}
%{\gamma _k} = \frac{{{p_k}\mathbb{E}\left\{ {{{\left| {\sum\limits_{m \in {{{\cal C}}_k}} {\hat g_{mk}^ * {g_{mk}}} } \right|}^2}} \right\}}}{{\sum\limits_{k' \ne k} {{p_{k'}}}\mathbb{E} \left\{ {{{\left| {\sum\limits_{m \in {{{\cal C}}_k}} {\hat g_{mk}^ * {g_{mk'}}} } \right|}^2}} \right\} + \sigma _m^2\mathbb{E}\left\{ {\sum\limits_{m \in {{{\cal C}}_k}} {{{\left| {{{\hat g}_{mk}}} \right|}^2}} } \right\}}}
%\end{equation}
Given the system bandwidth $W$, the uplink rate of user $k$ is given as  ${R_k} = W{\log _2}\left( {1 + {\gamma _k}}\right)$ by Shannon’s theory.  
\subsection{Parallel Computation Model}
We assume each user $k$ has a computationally intensive task of ${{\cal T}_{k}}\left( t \right)$ bits at the beginning of every discrete time step $t$, whose duration $\Delta t={{\tau }_{c}}$. The task sizes in different time steps are independent and uniformly distributed  over $\left[ {{\cal T}_{\min ,}}\,{{\cal T}_{\max ,}} \right]$, for every user $k \in {{\cal K}}$. Let $t_{k}^{d}$ denote the maximum tolerable delay of user $k$ to complete execution of the given task. Considering the delay constraint and energy consumption, the $k$-th user processes ${\cal T}_{k}^{local}\left( t \right)$ bits locally while offloading the remaining  ${\cal T}_{k}^{offload}\left( t \right)$ bits to the edge server at the CPU. 
\subsubsection{\bfseries Local computation}
Let $f_{k}^{local}\left( t \right)={{\alpha }_{k}}\left( t \right)f_{k}^{\max }$, for ${{\alpha }_{k}}\in \left[ 0,1 \right]$,  denote the $k$-th user processor speed (in cycle per second) allocated for local task execution. The entire task is offloaded to the edge server if ${{\alpha }_{k}}=0$, while ${{\alpha }_{k}}=1$ implies that the whole local processing capability is fully utilized and the remaining task bits are offloaded to the edge server. Accordingly, the proportion of locally computed task size at time step $t$ is expressed as ${\cal T}_{k}^{local}\left( t \right)=\min \left( {{\cal T}_{k}}\left( t \right),\frac{t_{k}^{d}f_{k}^{local}\left( t \right)}{{{N}_{cpb}}} \right)$, where ${{N}_{cpb}}$ corresponds to the number of cycles required to process one-bit task. Then, the time taken for local execution at time step $t$ is expressed as
\begin{equation}
t_k^{local}\left( t \right) = \min \left( {\frac{{{\cal T}_k^{local}\left( t \right){N_{cpb}}}}{{f_k^{local}\left( t \right)}},t_k^d} \right). 
\end{equation}
The corresponding energy consumption is given by
\begin{equation}
E_k^{local}\left( t \right) = \varsigma {\cal T}_k^{local}\left( t \right){N_{cpb}}{\left( {f_k^{local}\left( t \right)} \right)^2},
\end{equation}
where $\varsigma $  corresponds to the effective switched capacitance. 
\subsubsection{\bfseries Computation Offloading}
The remaining ${\cal T}_{k}^{offload}\left( t \right)=\max \left( 0,\,\,{{\cal T}_{k}}\left( t \right)-{\cal T}_{k}^{local}\left( t \right) \right)$ bits are offloaded to the edge server at the CPU for parallel computation. While offloading computation to the edge server, the experienced latency can be broken down into transmission delay, computing delay at the edge server, and transmission delay for retrieving the processed result. Assuming the retrieved data size after computation is much smaller compared to the offloaded data size, we ignore the retrieving delay. Then, the transmission delay to offload ${\cal T}_{k}^{offload}\left( t \right)$ data bits is expressed as  $t_k^{tr}\left( t \right) = \frac{{{{\cal T}}_k^{offload}\left( t \right)}}{{{R_k}\left( t \right)}}$,  where ${R_k}\left( t \right)$ is the uplink rate of user $k$ at time step $t$. The computing time at the edge server is given by $t_k^{comp}\left( t \right) = \frac{{{{\cal T}}_k^{offload}\left( t \right){N_{cpb}}}}{{f_k^{CPU}\left( t \right)}}$, where $f_{k}^{CPU}\left( t \right)$ is computation resource at the CPU allocated for user $k$,  which is in proportion to ${\cal T}_{k}^{offload}\left( t \right)$. The total offloading delay for edge computation is then given as  
\begin{equation}
t_k^{offload}\left( t \right) = t_k^{comp}\left( t \right) + t_k^{tr}\left( t \right).
\end{equation} 
The corresponding energy consumption for offloading computation is then expressed as
\begin{equation}
E_k^{offload}\left( t \right) = {p_k}\left( t \right)t_k^{tr}\left( t \right), 
\end{equation}
where ${p_k}\left( t \right) = {\eta _k}\left( t \right)\,p_k^{\max },\forall {\eta _k} \in \left[ {0,1} \right]$.
Then, the overall experienced latency ${{t}_{k}}\left( t \right)$ by user $k$ to execute ${{\cal T}_{k}}$ bits locally and at the edge is given as:
\begin{equation}
{t_k}\left( t \right) = \max (t_k^{local}\left( t \right),t_k^{offload}\left( t \right)).
\end{equation}
Similarly, the energy consumption ${{E}_{k}}\left( t \right)$ of user $k$ at time step $t$ can be expressed as
\begin{equation}
{E_k}\left( t \right) = E_k^{local}\left( t \right) + E_k^{offload}\left( t \right).
\end{equation}
\subsection{JCCRA Problem Formulation}
\vspace{-2pt}
According to the aforementioned communication and parallel computation models, we intend to jointly optimize the local processor speed $f_k^{local}\left( t \right)$ and uplink transmission power ${p_k\left( t \right)}$ for every user $k \in {{\cal K}}$ at each time step $t$, in order to minimize the total energy consumption of all users subject to the respective user-specific delay requirements. The JCCRA problem can mathematically be formulated to jointly determine $\left( {{\alpha _k}\left( t \right),{\eta _k}\left( t \right)} \right),\forall k \in {{\cal K}}$ as follows: 
\begin{equation}
\begin{array}{l}
	\mathop {\min }\limits_{\{ {\alpha _k}(t),{\eta _k}(t)|\forall k\} } \,\,\,\,\,\,\,\,\,\,\sum\limits_{k = 1}^K {{E_k}\left( t \right)} \\
	{\rm{\,\,\,\,subject}}\,{\rm{to}}\,\,\,\,\,\,\,\,\,\,\,\,\,{t_k}\left( t \right) \le t_k^d,\,\,\,\,\forall k\\
	\,\,\,\,\,\,\,\,\,\,\,\,\,\,\,\,\,\,\,\,\,\,\,\,\,\,\,\,\,\,\,\,\,\,\,\,0 \le {\alpha _k}\left( t \right) \le 1,\,\,\forall k\\
	\,\,\,\,\,\,\,\,\,\,\,\,\,\,\,\,\,\,\,\,\,\,\,\,\,\,\,\,\,\,\,\,\,\,\,\,\,0 \le {\eta _k}\left( t \right) \le 1,\,\,\,\forall k.
\end{array}
\end{equation}
(12) is a stochastic optimization problem in which the objective function and the first condition involve constantly changing random variables, i.e., task size and channel conditions. Consequently, the problem must be solved frequently, i.e., at every time step $t$, and the solution must converge rapidly within the ultra-low delay constraint. Therefore, it is challenging to solve the problem using iterative optimization algorithms with reasonable complexity. Furthermore, relying on a one-shot optimization, it is difficult to guarantee a steady long-term performance with conventional algorithms since they cannot adapt to the changes in the mobile edge network. 
%\begin{figure}[!t]       %   !t  -> to take it to the top
%\centering
%
%\includegraphics[width=2.5in,height=1.5in]{Fig1__}
%\includegraphics[width=3.34in,height=2.6in]{Fig_2_MARL}
%
%\includegraphics[width=3.4in,height=2.5in]{MARL_updated}
% where an .eps filename suffix will be assumed under latex, 
% and a .pdf suffix will be assumed for pdflatex; or what has been declared
% via \DeclareGraphicsExtensions.
%\vspace{-1mm}
%\caption{Multi-agent reinforcement learning framework: \emph{Illustration}}
%\vspace{-5mm}
%\end{figure}

\section{The Proposed Multi-Agent Reinforcement Learning-based Distributed JCCRA }  
\vspace{-2pt}
To cope up with the dynamics in the MEC while supporting efficient and adaptive joint resource allocation for every user, we propose a fully distributed JCCRA based on cooperative multi-agent reinforcement learning (MARL). We note that solving the JCCRA problem centrally at the CPU might allow for efficient management of the available radio and computing resources due to globally processing of network-wide information. However, the associated signaling and communication overheads are forbiddingly significant, especially given the ultra-low delay constraints. Accordingly, we transform the JCCRA problem into a partially observable cooperative Markov game that consists of $K$ learning agents. One of the main challenges in multi-agent domain is the non-stationarity of the environment since the agents update their policies concurrently, which might lead to learning instability.  To deal with this challenge, the agents are trained with the state-of-the-art multi-agent deep deterministic policy gradient (MADDPG) algorithm under the framework of centralized training and decentralized execution [8].
\subsection{DRL Formulation for JCCRA}
Each user is implemented as a DRL agent that learn to map local observation of the environment to efficient JCCRA actions by sequentially interacting with the mobile edge network in discrete time steps. Let ${{{\cal O}}_k}$, and ${{\cal S}}$ denote the local observation space of agent $k \in {{\cal K}}$, and the complete environment state space, respectively. At the beginning of each step $t$, agent $k$ perceives the local observation of the environment state ${o_k}\left( t \right):{{\cal S}} \mapsto {{{\cal O}}_k}$, and takes an action  ${a_k}\left( t \right) \in {{{\cal A}}_k}$ from its action space ${{{\cal A}}_k}$ according to the current JCCRA policy ${{\mu }_{k}}$. The interaction with the environment produces the next observation of the agent ${o_k}\left( {t + 1} \right) \in {{{\cal O}}_k}$ and a real-valued scalar reward ${r_k}\left( t \right):{{\cal S}} \times {{{\cal A}}_k} \mapsto \mathsf{\mathbb{R}}$ that guides the agent to constantly improve the policy until it converges. The goal of the agent is, therefore, to derive an optimal JCCRA policy $\mu _{k}^{*}$ that maximizes the expected long-term discounted cumulative reward, defined as ${{J}_{k}}\left( {{\mu }_{k}} \right)=\mathbb{E}\left[ \sum\limits_{t=1}^{T}{{{\varepsilon }^{t-1}}{{r}_{k}}}\left( t \right) \right]$, where $\varepsilon \in \left[ 0,\,1 \right]$ is a discount factor and  $T$ is the total number of time steps (horizon). 
In the sequel, we define the \emph{local observation}, \emph{action}, and \emph{reward} of agent $k$  for the JCRRA at a given time $t$.
\subsubsection{ Local observation}
At the beginning of each time step $t$, the local observation  of agent $k$ is defined as ${{o}_{k}}\left( t \right)\triangleq \left[ {{\cal T}_{k}}(t),t_{k}^{d},{{R}_{k}}(t-1) \right]$, where ${{\cal T}_{k}}(t)$, $t_{k}^{d}$, , and ${{R}_{k}}(t-1)$ are the incoming task size, user-specific application deadline, and the uplink rate from the previous time step, respectively. Note that the local observation is only a partial view of the environment, i.e., it does not include information about the other agents.  
\subsubsection{ Action}
Based on the local observation of the environment, the agent takes action ${a_k}\left( t \right) \buildrel \Delta \over = \left[ {{\alpha _k}\left( t \right),{\eta _k}\left( t \right)} \right]$, i.e., local processor speed
$f_k^{local}\left( t \right) = {\alpha _k}\left( t \right)f_k^{\max }$, and uplink transmit power ${p_k}\left( t \right) = {\eta _k}\left( t \right)\,p_k^{\max }$. 
\subsubsection{ Reward}
To reflect the design objective of the JCCRA problem and enforce a cooperative behavior among the agents, we define joint reward as 
\begin{equation}
{r_k}(t) =  - \sum\limits_{k = 1}^K {{\xi _k}{E_k}\left( t \right),\,\,} \forall k \in {{\cal K}},
\end{equation}
where ${\xi _k} = 1$ if ${t_k}\left( t \right) \le t_k^d$, otherwise ${\xi _k} = 10$ to punish potentially selfish behavior of the agents that leads to failure in meeting the delay constraint. 
\subsection{MADDPG Algorithm for Distributed JCCRA}
During the offline training phase, the agents can be trained in a central unit, which presents an opportunity to share extra information among the agents for easing and accelerating the training process. Specifically, agent $k$ has now access for the joint action $a\left( t \right)=\left( {{a}_{1}}\left( t \right),...,{{a}_{K}}\left( t \right) \right)$, and full observation of the environment state ${s_k}\left( t \right) = \left( {{o_k}\left( t \right),{o_{ - k}}\left( t \right)} \right)$, where ${o_{ - k}}\left( t \right)$ is the local observation of other agents at time step $t$. The extra information endowed to the agents, however, is discarded during the execution phase, meaning the agents rely on their local observation entirely to make JCCRA decisions in a fully distributed manner.   

As an extension of the single-agent deep deterministic policy gradient (DDPG) algorithm [9], the MADDPG is an actor-critic policy gradient algorithm. Specifically, the MADDPG agent $k$ employs two main neural networks: actor network with parameters $\theta _{k}^{\mu }$ to approximate a JCCRA policy ${{\mu }_{k}}\left( {{o}_{k}}|\theta _{k}^{\mu } \right)$ and a critic network, parametrized by $\theta _{k}^{Q}$, to approximate a state-value function ${{Q}_{k}}\left( {{s}_{k}},a|\theta _{k}^{Q} \right)$, along with their respective time-delayed copies, $\theta _{k}^{{{\mu }'}}$ and $\theta _{k}^{{{\mu }'}}$, which serve as targets. At time step $t$, the actor deterministically maps local observation ${{o}_{k}}\left( t \right)$ to a specific continuous action ${{\mu }_{k}}\left( {{o}_{k}}\left( t \right)|\theta _{k}^{\mu } \right)$ and then, a random noise process ${{{\cal N}}_k}$ is added to generate an exploratory policy such that ${{a}_{k}}\left( t \right)={{\mu }_{k}}\left( {{o}_{k}}\left( t \right)|\theta _{k}^{\mu } \right)+ {{{\cal N}}_k}\left( t \right)$. The environment, which is shared among the agents, collects the joint action $a\left( t \right) = \left\{ {{a_k}\left( t \right),\forall k \in {{\cal K}}} \right\}$ and returns the immediate reward ${{r}_{k}}\left( t \right)$ and the next observation ${{o}_{k}}\left( t+1 \right)$ to the respective agents. To make use of the experience in later decision-making steps efficiently and improve the stability of the training, the agent’s transition along with the extra information ${{e}_{k}}\left( t \right)=\left( {{s}_{k}}\left( t \right),a\left( t \right),{{r}_{k}}\left( t \right),{{s}_{k}}\left( t+1 \right) \right)$ is saved in the replay buffer ${\cal D}_k$ of the agent.

To train the main networks, a mini batch of $B$ samples $\left. {\left( {s_k^i,a_{}^i,r_k^i,s_k^{i + 1}} \right)} \right|_{i = 1}^B$, is randomly drawn from the replay buffer ${\cal D}_k$, where $i$ denotes a sample index. The critic network is updated to minimize the following loss function:
\begin{equation}
	{{\cal L}_k}\left( {\theta _k^Q} \right) = \frac{1}{B}\,\,\sum\nolimits_i {\,{{\left( {y_k^i - {Q_k}\left( {s_k^i,{a^i}|\theta _k^Q} \right)} \right)}^2}} \,\,,
\end{equation}
where $y_{k}^{i}$ is the target value expressed as $y_{k}^{i}={{\left. r_{k}^{i}+\varepsilon \,\,{{{{Q}'}}_{k}}\left( s_{k}^{i+1},{{a}_{i+1}}|\theta _{k}^{Q'} \right) \right|}_{{{a}_{^{i+1}}}=\left\{ {{{{\mu }'}}_{k}}\left( o_{k}^{i+1} \right),\forall k\in \cal{K} \right\}}}$. On the other hand, the parameters of the actor network are updated along the deterministic policy gradient given as   
\begin{multline}
	{\nabla _{\theta _k^\mu }}J\left( {{\mu _k}|\theta _k^\mu } \right) \approx \frac{1}{B}\,\Biggl[  \sum\nolimits_i {{\nabla _{{a_k}}}} {Q_k}\left( {s_k^i,{a_i}|\theta _k^Q} \right)
	\times  \\\nabla_{\theta _k^\mu }\,{\mu _k}\left( {o_k^i|\theta _k^\mu } \right) \Biggl]\Biggl|_{{a_i} = \left\{ {{\mu _k}\left( {o_k^i} \right),\,\forall k \in {{\cal K}}} \right\}} .
\end{multline}           
The target parameters in both actor and critic networks are updated as $\theta _k^{\mu '} \leftarrow \tau \theta _k^\mu  + \left( {1 - \tau } \right)\theta _k^{\mu '}$, and $\theta _k^{Q'} \leftarrow \tau \theta _k^Q + \left( {1 - \tau } \right)\theta _k^{Q'}$, respectively, where $\tau $ is a constant close to zero. 

It is worth mentioning that the proposed JCCRA scheme is trained offline. Once it converges, real-time decision making can be conducted in a fully distributed fashion by simply performing inference with the trained actor network of each agent, relying on the respective local observation only. 
\section{Performance Analysis }
We consider $M=100$ APs and $K=10$ users uniformly distributed at random over an area of 1 km$^{2}$. The APs are connected to the CPU via ideal fronthaul links. We assume only 30\% of the entire APs are clustered to serve each user, i.e. ${N_k} = 0.3M,\forall k \in {{\cal K}}$. The system bandwidth is set to 5MHz and all active users share this bandwidth without channelization. Denoting the standard deviation of the shadow fading by ${{\sigma }_{sh}}$, the large-scale channel gain is given as ${{\beta }_{mk}}={{10}^{\frac{P{{L}_{mk}}}{10}}}{{10}^{\frac{{{\sigma }_{sh}}{{z}_{mk}}}{10}}}$, where ${z_{mk}}\sim {{\cal N}}\left( {0,1} \right)$. The path loss $P{{L}_{mk}}$ is given according to the three-slope model [10] as follows: 
\begin{equation}
  P{L_{mk}} =
    \begin{cases}
      - L - 35{\log _{10}}\left( {{d_{mk}}} \right) & \text{if 
${d_{mk}} > d_1$}\\
      - L - 10{\log _{10}}\left( {d_{mk}^2\,d_1^{1.5}} \right) & \text{if ${d_0} < {d_{mk}} \le {d_1}$}\\
      - L - 10{\log _{10}}\left( {d_0^2\,d_1^{1.5}} \right) & \text{if ${d_{mk}} \le {d_0}$}
    \end{cases},      
\end{equation}
where ${d_{mk}}$  is the distance between the user $k$ and AP $m$, and $L$ is 
\begin{multline}
L = 46.3 + 33.9\,{\log _{10}}\left( f \right) - 13.82\,{\log _{10}}\left( {{h_{AP}}} \right) \\- \left( {1.1\,{{\log }_{10}}\left( f \right) - 0.7} \right){h_u} +  {1.56\,{{\log }_{10}}\left( f \right) - 0.8} \,.
\end{multline}
Here,  $f$  is the carrier frequency (in MHz), ${{h}_{AP}}$ is the AP antenna height (in m), and  ${{h}_{u}}$ denotes the user antenna height (in m). Further, no shadowing is considered unless  ${{d}_{mk}}>{{d}_{1}}$. The edge server at the CPU has computation capacity ${{f}^{CPU}}=100\,\text{GHz}$, while the users are equipped with local processor with $f_{k}^{\max }=1\,\text{GHz}$. The length of a time step $\Delta t$ is set to 1ms, and each user generates a task of a random size that is uniformly distributed in the range of 2.5 to 7.5 kbits at the beginning of every time step $t$. Furthermore, Furthermore, additional simulation parameters are set to  ${{N}_{cpb}}=500$, $\varsigma ={{10}^{-27}}$, $p_{k}^{\max }=0.1 \text{W}$, ${{\sigma }_{sh}}=10\text{dB}$, and $t_{k}^{d}=\Delta t={{\tau }_{c}}=1 \text{ms}$.

For the proposed MADDPG-based JCCRA implementation, the actor and critic networks of each agent are implemented with fully connected neural networks, which consist of  128, 64, and 64 nodes.  All the hidden layers are activated by ReLu function, while the outputs of the actor are activated by sigmoid. The parameters of critic and actor networks are updated with adaptive moment (Adam) optimizer with the learning rates of 0.001 and 0.0001. Further, we set the discount factor ${{\gamma }_{k}}=0.99$, target update parameter $\tau =0.005$, and mini batch size $B=128$. In our simulations, the agents are trained for several episodes, each consisting of 100 steps.

To evaluate the performance of the proposed MADDPG-based JCCRA, we compare it with three benchmarks: 
\subsubsection{\bfseries Centralized DDPG-based JCCRA scheme}
This approach refers to a DDPG-based centralized resource allocation scheme implemented at the CPU. We adopt the same neural network structure and other hyperparameters as the MADDPG-based scheme to train the actor and critic in the single DDPG agent. Since global information of the entire network is processed centrally at the CPU to make JCCRA decisions for all users, we can potentially obtain the most efficient resource allocation. Consequently, this baseline serves as a target performance benchmark. However, as it involves significant signaling and communication overheads, this scheme might be infeasible to support time-sensitive applications.
\begin{figure}[!t]       %   !t  -> to take it to the top
%\vspace{-16pt}
\vspace{-13pt}
\centering
\includegraphics[width=3.20in,height=2.2in]{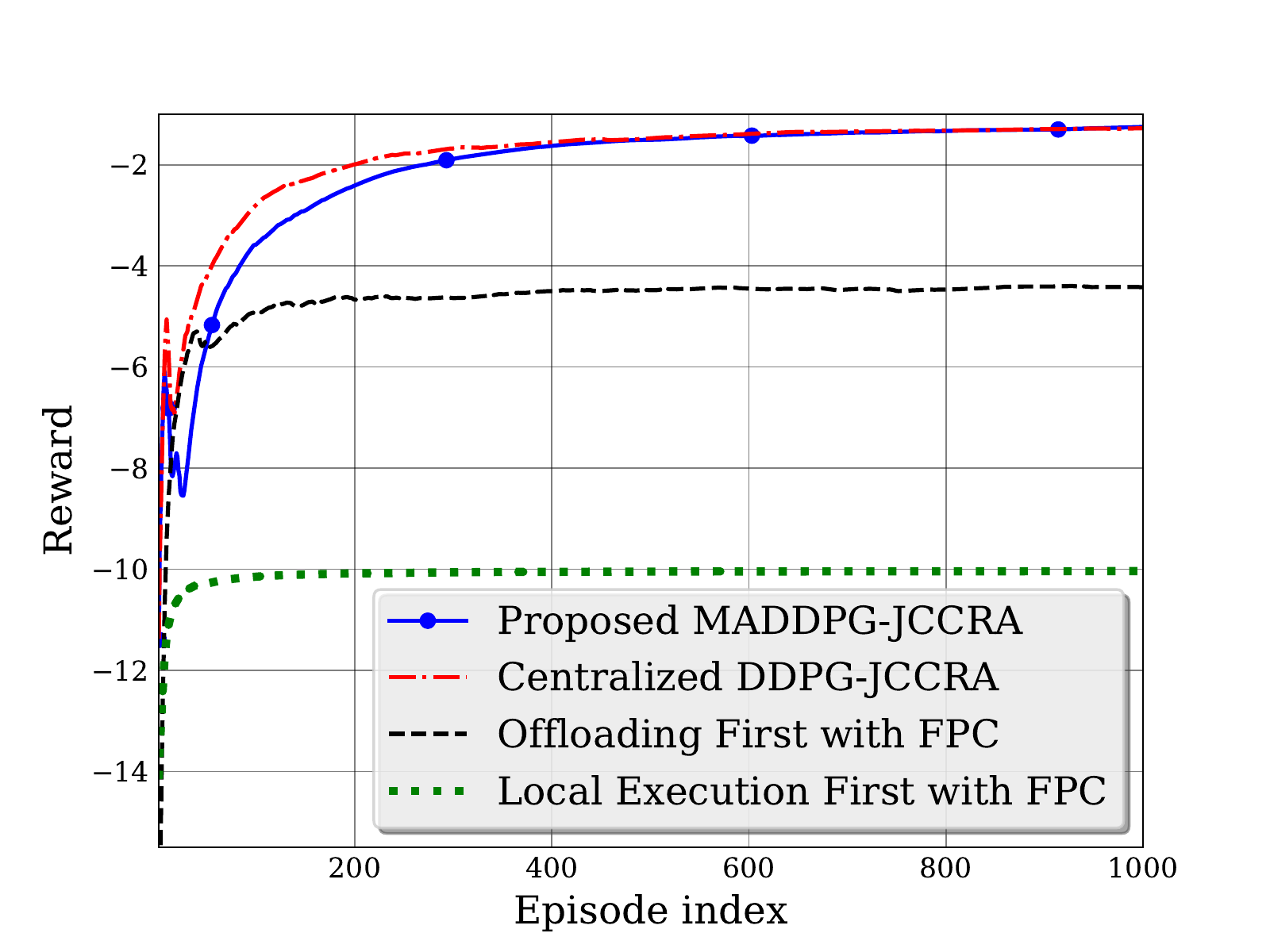}
% where an .eps filename suffix will be assumed under latex, 
% and a .pdf suffix will be assumed for pdflatex; or what has been declared
% via \DeclareGraphicsExtensions.
\vspace{-3pt}
\caption{ Average reward with training process }
%\vspace{-15pt}
\end{figure}
\begin{figure}[!h]    %   !t  -> to take it to the top
\centering
\vspace{-10pt}
\includegraphics[width=3.20in,height=2.2in]{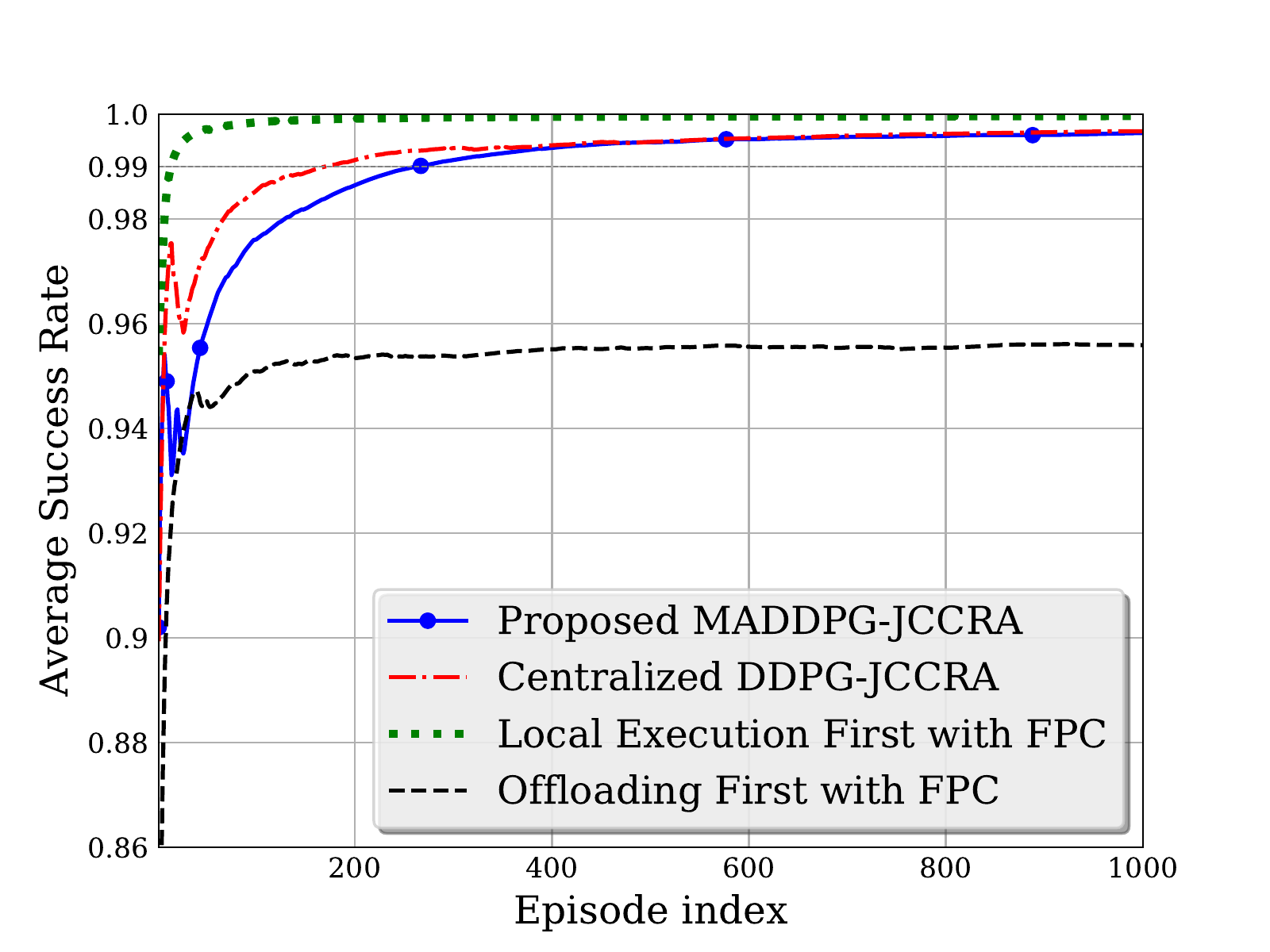}
%\includegraphics[width=0.8\linewidth]{Fig1_up}
% where an .eps filename suffix will be assumed under latex, 
% and a .pdf suffix will be assumed for pdflatex; or what has been declared
% via \DeclareGraphicsExtensions.
\vspace{-3pt}
\caption{Average success rate with training process }
\vspace{-13pt}
\end{figure}
\subsubsection{\bfseries Offloading-first with an uplink fractional power control (FPC) scheme}
This approach preferably offloads the computation to the edge server with the aim of aggressively exploiting the reliable access link provided by cell-free massive MIMO while saving the local processing energy consumption. The uplink transmit power for the $k$-th user is determined by the standard fractional power control (FPC) [11] as follows: 
\begin{equation}
{p_k} = \min \left( {p_k^{\max },{p_0}\lambda _k^{ - \nu }} \right),
\end{equation}
where ${p_0} =  - 35\,{\rm{dBm}}$, ${\lambda _k} = \sum\limits_{m = 1}^{{N_k}} {{\beta _{mk}}} $, and $\nu  = 0.5$. 
%\vspace{2mm}

\subsubsection{\bfseries Local execution-first with an uplink fractional power control (FPC) scheme}
The entire local processing capability is fully utilized, i.e., $f_k^{local} = f_k^{\max }$, and the remaining task bits are offloaded to the edge server with uplink transmit power given according to (18).

In Fig. 2, we investigate the convergence and performance of the proposed MADDPG-based JCCRA scheme by evaluating the total reward accumulated during the training process. It can be observed that the performance of the proposed distributed scheme has converged to the target performance benchmark, i.e. the centralized DDPG-based JCCRA, while relieving the large overhead of the latter. This implies that by relying on local observation and additional information provided during the offline centralized training phase, the agents manage to learn efficient joint resource allocation. Moreover, the proposed scheme has significantly outperformed the heuristic baselines in terms of total reward accumulated.

In Fig. 3, we compare the average success rate of the users in attaining the delay constraint of the time-sensitive applications.  One can observe that the proposed MADDPG-based JCCRA has achieved more than 99\% success rate in accomplishing the tasks within the deadline as the training episodes progress. Even though \emph{local execution-first} policy achieves 100\% success rate, the performance comes at the cost of high energy consumption, as reflected in Fig. 2. Moreover, the fact that the policy of the learning agents outperforms the \emph{offloading-first} baseline implies even if cell-free massive MIMO can provide reliable access links to the users, aggressive computation offloading can result in performance degradation due to ineffective use of the limited resources. 

\begin{figure}[!t]
\centering
\vspace{-13pt}
%\vspace{-18pt}
    \begin{subfigure}[!t]{\linewidth}                %  {0.3\textwidth}
    \centering
    \includegraphics[width=3.20in,height=2.2in]{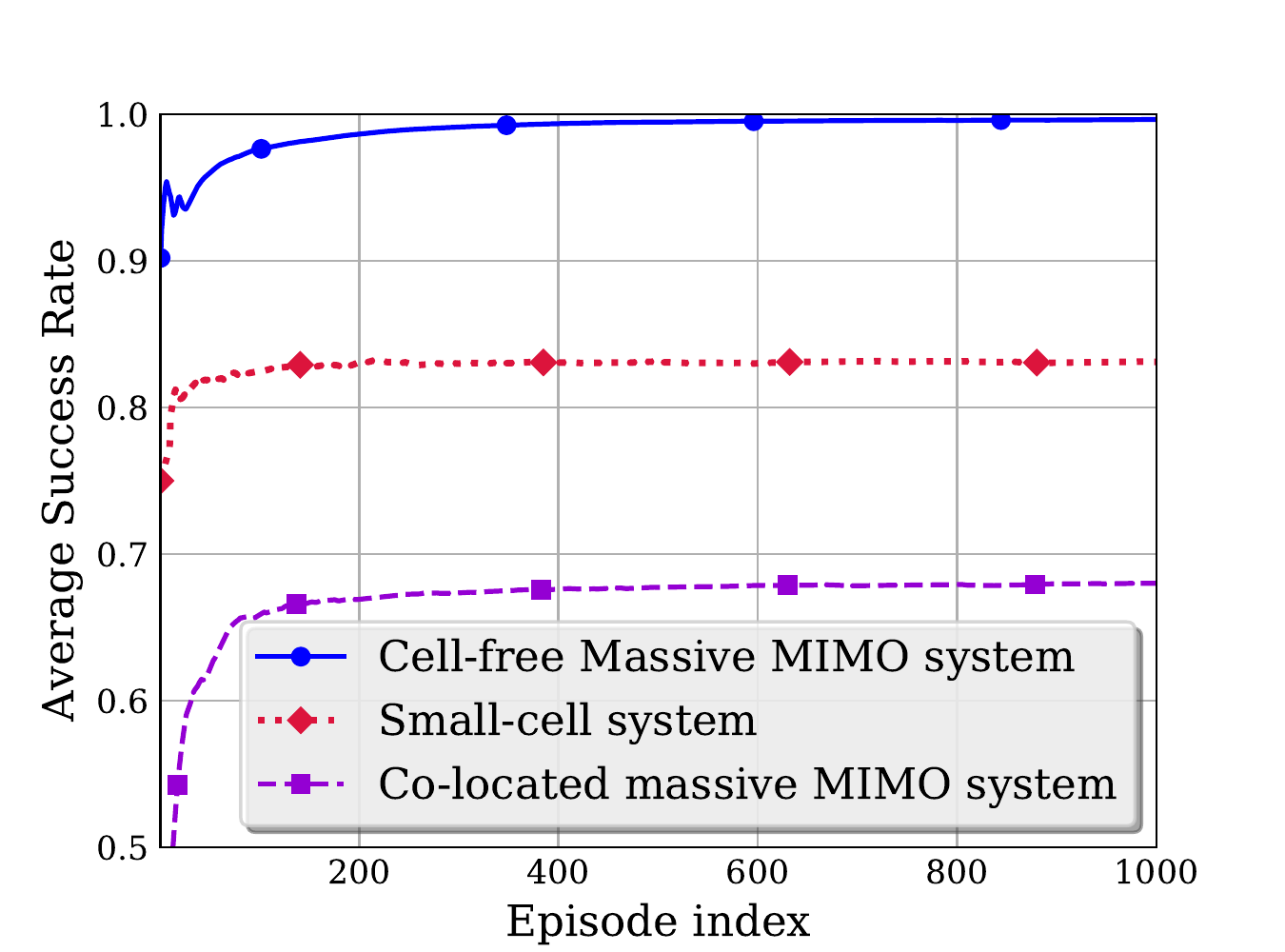}
        %\vspace{-3pt}
    \vspace{-2.0pt}
    \caption{Average success rate} \label{fig1}
    %\vspace{-15pt}
\end{subfigure} %\hfill
\begin{subfigure}[!t]{\linewidth}
    \centering
  %  \vspace{-1.5pt}
    %\vspace{-1.0pt}
  %  \includegraphics[width=0.9\linewidth, height=1.80in]{Fig_Mini_Globecom_up_rew}
    %\includegraphics[width=3.0in,height=1.95in]{Fig_Mini_Globecom_up_rew}
     \includegraphics[width=3.20in,height=2.2in]{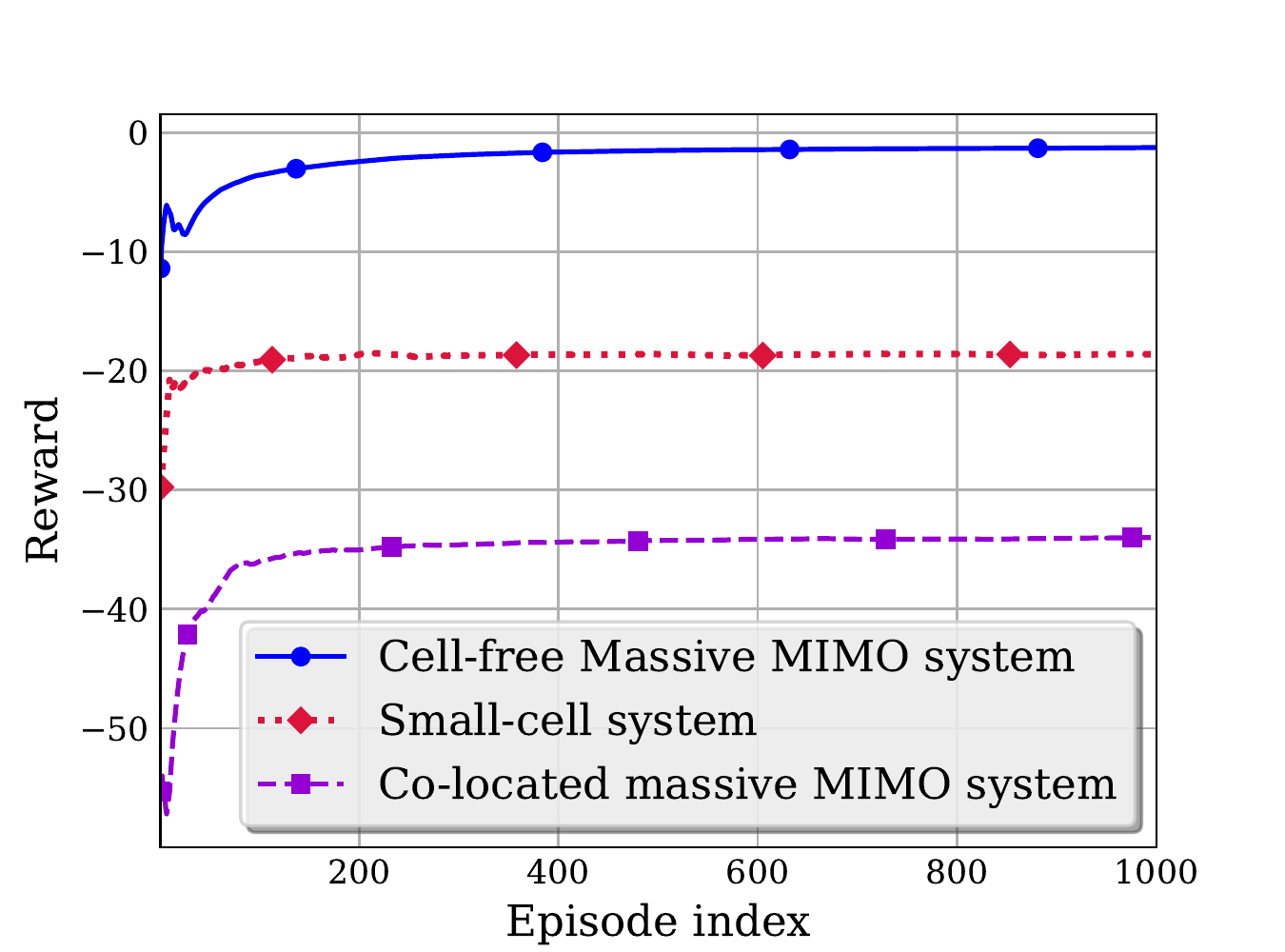}
        %\vspace{-3pt}
        \vspace{-2pt}
    \caption{Reward} \label{fig2}
    %\vspace{-15pt}
\end{subfigure}        %\hfill

\vspace{-3pt}
\caption{Performance comparison of the proposed algorithm over cell-free and cellular MEC architectures}
\vspace{-5pt}
\label{fig:mylabel}
%\vspace{-15pt}
\vspace{-10pt}
\end{figure}

We then implemented the proposed MADDPG-based JCCRA scheme in two different cellular MEC architectures, including a small-cell system and a single-cell system with co-located massive MIMO, to further investigate the gain obtained from cell-free architecture. For the small-cell MEC system, each user is served by a single AP with the largest large-scale fading coefficient according to the methodology in [5], whereas in co-located massive MIMO case, each user connects to a base station with total number of  ${N_k}$ antennas. From Fig.’s 4(a) and 4(b), we can observe that the proposed algorithm over a cell-free system has significantly outperformed the cellular MEC systems, in terms of average success rate and total reward accumulated. This is attributed to the channel quality degradation due to spatial co-channel interference from other near-by cells in the small-cell system case, while in the single-cell MEC system, users are subjected to large interference from all co-located antennas under MRC, hence limiting the uplink rate for computation offloading.

%\vspace{-2pt}

\section{Conclusion and future works }
%\vspace{-2pt}
In this paper, we presented a novel distributed joint communication and computing resource allocation (JCCRA) scheme based on cooperative multi-agent reinforcement learning framework for the cell-free massive MIMO-enabled mobile edge network. More specifically, each trained agent can make intelligent and adaptive joint resource allocation in real-time, based on local observation only, in order to minimize the total energy consumption while meeting the respective delay constraints of the dynamic MEC system. The performance of the proposed distributed JCCRA scheme is shown to converge to the centralized target baseline, without resorting to large overhead. Finally, we have shown that the proposed algorithm in cell-free system has outperformed two cellular MEC systems by wide margins. In the future, it would be interesting to extend the current formulation to determine a set of APs that are adaptive to dynamic user mobility.  
%\vspace{-15pt}
%The simulation results indicate that despite the difficulties posed by the dynamic MEC environment, the %agents learn adaptive and efficient JCCRA policies. 
%Furthermore, user scheduling should be incorporated to capture the dynamic traffic load among all users. 
%\vspace{-5pt}

%\vspace*{-10pt}
%\vspace*{-4pt}
\section*{Acknowledgment}
%\vspace*{-4pt}
This work was supported by the National Research Foundation of
Korea (NRF) grant funded by the Korea government (MSIT)
(No.2020R1A2C100998413).
%\footnotesize{This work was supported by the National Research Foundation of
%Korea (NRF) grant funded by the Korea government (MSIT)
%(No.2020R1A2C100998413).}
%\footnotesize{This work was supported by Institute of Information & Communications Technology Planning & Evaluation(IITP) grant funded by the Korea government(MSIT) (No.2021-0-00467, Intelligent 6G Wireless Access System)}.
%\vspace*{-5pt}


\begin{thebibliography}{15}
%\vspace*{-2pt}
\bibitem{IEEEhowto:kopka}
J. Li, H. Gao, T. Lv, and Y. Lu, "Deep reinforcement learning based computation offloading and resource allocation for MEC," 2018 \emph{IEEE Wireless Communications and Networking Conference (WCNC)}, Barcelona, 2018.\hskip 1em plus
0.5em minus 0.4em\relax

\bibitem{IEEEhowto:kopka}
L. Huang, X. Feng, C. Zhang, L. Qian, and Y. Wu, “Deep reinforcement learning-based joint task offloading and bandwidth allocation for multi-user mobile edge computing,” \emph{Digit. Commun. Netw.}, vol. 5, no. 1, pp. 10–17, 2019.\hskip 1em plus
0.5em minus 0.4em\relax  

\bibitem{IEEEhowto:kopka}
Y. Dai, K. Zhang, S. Maharjan, and Y. Zhang, "Edge Intelligence for Energy-Efficient Computation Offloading and Resource Allocation in 5G Beyond," in \emph{IEEE Transactions on Vehicular Technology}, vol. 69, no. 10, pp. 12175-12186, Oct. 2020.\hskip 1em plus
0.5em minus 0.4em\relax 

\bibitem{IEEEhowto:kopka}
Chen Z. and Wang, X. “Decentralized computation offloading for multi-user mobile edge computing: a deep reinforcement learning approach”. \emph{J Wireless Com Network} 2020, 188 (2020).  \hskip 1em plus
0.5em minus 0.4em\relax  
 
\bibitem{IEEEhowto:kopka}
H. Q. Ngo, A. Ashikhmin, H. Yang, E. G. Larsson and T. L. Marzetta, "Cell-Free Massive MIMO Versus Small Cells," in \emph{IEEE Transactions on Wireless Communications}, vol. 16, no. 3, pp. 1834-1850, March 2017. \hskip 1em plus
0.5em minus 0.4em\relax   
  
\bibitem{IEEEhowto:kopka}
S. Mukherjee and J. Lee, "Edge Computing-Enabled Cell-Free Massive MIMO Systems," in \emph{IEEE Transactions on Wireless Communications}, vol. 19, no. 4, pp. 2884-2899, April 2020. \hskip 1em plus
0.5em minus 0.4em\relax  

\bibitem{IEEEhowto:kopka}
M. Ke, Z. Gao, Y. Wu, X. Gao, and K. -K. Wong, "Massive Access in Cell-Free Massive MIMO-Based Internet of Things: Cloud Computing and Edge Computing Paradigms," in \emph{IEEE Journal on Selected Areas in Communications}, vol. 39, no. 3, pp. 756-772, March 2021. \hskip 1em plus
0.5em minus 0.4em\relax 

\bibitem{IEEEhowto:kopka}
R. Lowe et al., “Multi-Agent Actor-Critic for Mixed Cooperative-Competitive Environments”, arXiv preprint arXiv:1706.02275, 2020.  \hskip 1em plus
0.5em minus 0.4em\relax

\bibitem{IEEEhowto:kopka}
T. P. Lillicrap, J. J. Hunt, A. Pritzel, N. Heess, T. Erez, Y. Tassa, D. Silver, and D. Wierstra, “Continuous control with deep reinforcement learning,” in Proc. \emph{International Conference on Learning Representations (ICLR)}, 2016.  \hskip 1em plus
0.5em minus 0.4em\relax

%\bibitem{IEEEhowto:kopka}
%T. P. Lillicrap, et al., “Continuous control with deep reinforcement learning,” in Proc. %\emph{International Conference on Learning Representations (ICLR)}, 2016.  \hskip 1em %plus
%0.5em minus 0.4em\relax

\bibitem{IEEEhowto:kopka}
Ao Tang, JiXian Sun, and Ke Gong, "Mobile propagation loss with a low base station antenna for NLOS street microcells in urban area," \emph{IEEE VTS 53rd Vehicular Technology Conference}, Spring 2001. Proceedings (Cat. No.01CH37202), Rhodes, Greece, 2001, pp. 333-336 vol.1. \hskip 1em plus
0.5em minus 0.4em\relax 

\bibitem{IEEEhowto:kopka}
R. Nikbakht, R. Mosayebi, and A. Lozano, "Uplink Fractional Power Control and Downlink Power Allocation for Cell-Free Networks," in \emph{IEEE Wireless Communications Letters}, vol. 9, no. 6, pp. 774-777, June 2020. \hskip 1em plus
0.5em minus 0.4em\relax 


\end{thebibliography}
\end{document}